\documentclass[aps,preprint,amsmath,showpacs,amssymb,nofootinbib]{revtex4}
\usepackage{epstopdf}

\usepackage{graphicx}
\usepackage{color}
\usepackage{latexsym}

\usepackage{feynarts}

\newcommand{\beq}{\begin{equation}}
\newcommand{\eeq}{\end{equation}}
\newcommand{\beqa}{\begin{eqnarray}}
\newcommand{\eeqa}{\end{eqnarray}}
\newcommand{\ba}{\begin{array}}
\newcommand{\ea}{\end{array}}
\newcommand{\no}{\nonumber}

\def\delr            {\!\stackrel{\leftrightarrow}{\partial^\mu}\!}

\begin{document}

\title{ Lepton Flavor Violating  $l \to l' \gamma$ and $Z \to l \bar l'$ Decays Induced by Scalar Leptoquarks}

\author{Rachid Benbrik$^{1,2}$\footnote{Email:
rbenbrik@phys.cycu.edu.tw\\ Address after Aug. 2008: Department of
Physics, National Cheng Kung University, Tainan 701, Taiwan.}}

\author{Chun-Khiang Chua$^1$\footnote{Email:
ckchua@phys.cycu.edu.tw} }%
\affiliation{%
$^1$Department of Physics, Chung Yuan Christian University
  Chung-Li, Taiwan 320, Republic of China\\
$^2$LPHEA, D\'epartement de Physique,Facult\'e des Sciences-Semlalia B.P 2390 Marrakech, Morocco.
}%

\begin{abstract}
Motivated by the recent muon $g-2$ data, we study the lepton
flavor violating $l \to l' \gamma$ and $Z \to l \bar l'$ ($l, l' =
e, \mu, \tau$) decays with $l \neq l'$ in a scalar leptoquark
model. Leptoquarks can produce sizable LFV $l\to l'\gamma$ decay
rates that can be easily reached by present or near future
experiments.
Leptoquark masses and couplings are constrained by the muon $g-2$
data and the current $l \to l' \gamma$ bounds. We predict $Br(Z
\to \tau^\mp e^\pm)$ reaching the present limit ($10^{-5}$) and
$Br(Z \to \mu^\mp \tau^\pm)$ reaching $2\times 10^{-8}$, which
will be accessible by future linear colliders, whereas, the
current bounds on LFV impose very strong constraints on the $Br(Z
\to \mu^\mp e^\pm)$ and the ratio is too low to be observed in the
near future.
\end{abstract}


\maketitle

\section{Introduction}
\label{sec:introduction}

The excess value of the anomalous magnetic moment of muon was
reported by the E821 collaboration at BNL~\cite{Bennet}
\begin{equation}
a^{{\rm exp}}_\mu = 116\, 592\, 080(63) \times 10^{-11}.
\end{equation}
The Standard Model prediction for $a^{SM}_\mu$ with QED, hadronic
and electroweak contributions is~\cite{Miller,Aoyama:2007mn}
\begin{equation}
a^{{\rm SM}}_\mu = 116\, 591\, 785 (61) \times 10^{-11}.
\end{equation}
with the experimental value of $(g-2)/2$, the comparison gives
\begin{equation}
\Delta a_\mu \equiv a^{{\rm exp}}_\mu -a^{{\rm SM}}_\mu = (295\pm
87.7) \times 10^{-11},  \qquad (3.4 \sigma) \label{amue}
\end{equation}
The 3.4 standard deviation difference between the two, may be a
hint of new physics contribution.

It has been shown that contributions from leptoquark (LQ)
exchanges are capable to resolve the above
deviation~\cite{Bigi:1985jq, Mahanta:2001yc,Cheung:2001ip}. Leptoquarks are
vector or scalar particles carrying both lepton and baryon
numbers. LQs can be quite naturally introduced in the low-energy
theory as a relic of a more fundamental theory at some high-energy
scale, such as grand unified theories (GUT) \cite{ps,GUT}. In some
models, it is possible to have leptoquarks at TeV
scale~\cite{lowLQ}. The low-energy LQ phenomenology has received
considerable attention. Possible LQ manifestations in various
processes have been extensively investigated
\cite{lowLQ}-\cite{Benbrik:2008ik}. Various constraints on LQ
masses and couplings have been deduced from existing experimental
data and prospects for the forthcoming experiments have been
estimated. Direct searches of LQs as s-channel resonances in deep
inelastic ep-scattering and pair production in hadron colliders
placed lower limits on their mass $M_{LQ} \geq 73-298$GeV
\cite{pdg} depending on the LQ types and couplings. The interest
on leptoquarks has been renewed during the last few years since
ongoing collider experiments have good prospects for searching
these particles \cite{LHC}. For a recent review of leptoquarks,
one is referred to \cite{LQreview}.

Lepton Flavor Violation (LFV) are powerful tools to search for new
physics. The present experimental limits give~\cite{pdg}:
\beqa
\label{lfv1}
{\rm Br}(\mu \to e \gamma) &<& 1.2 \times 10^{-11}, \\
\label{lfv2}
{\rm Br}(\tau \to e \gamma) &<& 1.1 \times 10^{-7}, \\
\label{lfv3} {\rm Br}(\tau \to \mu \gamma) &<& 6.8 \times 10^{-8}.
\eeqa
Since effects of leptoquark interactions can manifest in $a_\mu$,
it is very likely that they can also give interesting
contributions to these $l\to l'\gamma$
processes~\cite{Mahanta:2001yc,Cheung:2001ip}. There are
considerable efforts on experiments that aim at pushing the
sensitivity of ${\rm Br}(\mu \to e \gamma)$ down by two order of
magnitudes~\cite{LQll'gamma}. B factories and the upgraded super B
factory can probe the $\tau\to e\gamma,\,\mu\gamma$ decays at
better sensitivities.

The $Z \to \ell \bar\ell'$ decays are among the LFV interactions
and the theoretical predictions of their branching ratios  in the
framework of the SM are extremely small \cite{Riemann, Ganapathi,
Illana}.
These results are far from the experimental limits obtained at
LEP1 \cite{pdg}:
\begin{eqnarray}
Br(Z\rightarrow e^{\pm} \mu^{\mp}) &<& 1.7\times 10^{-6} \,\,\,\,,
\\
Br(Z\rightarrow e^{\pm} \tau^{\mp}) &<& 9.8\times 10^{-6}\,\,\,\,,
\\
Br(Z\rightarrow \mu^{\pm} \tau^{\mp}) &<& 1.2\times 10^{-5} \,\,.
\label{Expr1}
\end{eqnarray}
Better sensitivities are expected from the Giga-Z modes at future
colliders, such as International Linear Collider (ILC), to have
\cite{AguilarSaavedra:2001rg,Heinemeyer:2007aw,Erler:2008ek}:
\begin{eqnarray}
Br(Z\rightarrow e^{\pm} \mu^{\mp}) &<& 2\times 10^{-9} \, , \\
Br(Z\rightarrow e^{\pm} \tau^{\mp}) &<& \kappa\times 6.5\times 10^{-8}
\, , \\
Br(Z\rightarrow \mu^{\pm} \tau^{\mp}) &<& \kappa\times 2.2\times
10^{-8}, \label{Expr2}
\end{eqnarray}
with $\kappa\simeq0.2-1.0$.
It will be interesting to study the leptoquark contributions to
the $Z\to l\bar l'$ processes.

The aim of the present paper is to study the leptoquark effects in
various LFV processes including $l\to l'\gamma$ and $Z\to l\bar
l'$ decays, while considering leptoquark contribution to $a_\mu$
as a solution to the muon anomalous moment discrepancy.
The layout of the present paper is as follows: In Sec. II we
introduce the formalism.  We then use it in Sec. III to study the
leptoquark contributions to $a_\mu$ and LFV processes including
$l\to l'\gamma$ and $Z\to l\bar l'$ decays. Sec. IV contains our
conclusions. Some formulas and low energy constraints are given in
Appendices.

\section{Formalism}
\subsection{Scalar Leptoquark Interactions}

In this section we list the relevant parts of the scalar
leptoquark Lagrangian. We consider isosinglet scalar leptoquarks.
The effective Lagrangian describing the leptoquark interactions in
the mass basis is given by \cite{CHH1999,Lagr1}:
\begin{eqnarray}
 {\mathcal{L}}_{LQ}
 &=&
 \label{lag}
 \overline{u^c_a} \bigg(h^{'}_{ai} \Gamma_{k,S_R} P_L+h_{ai}\Gamma_{k,S_L} P_R \bigg) e_i S^*_k
 +\overline{e_{j}} \bigg( h^{'*}_{aj} \Gamma^\dagger_{S_R,k} P_R+h^*_{aj} \Gamma^\dagger_{S_L,k} P_L
 \bigg) u^c_a S_k
 \\\no
 &-& e Q_{(u^c)} A_\mu \overline{u^c_a} \gamma^\mu u^c_a - ieQ_{S} A_\mu S^*_{k} \delr S_{k}
 + ieQ_{S} \tan\theta_W Z_\mu S^*_{k} \delr S_{k}
 \\\no
 &-& \frac{e }{s_W c_W} Z_\mu \overline{u^c_a}\gamma^\mu
 \bigg( (T_{3(u^c)} - Q_{(u^c)} s^2_W) P_R - Q_{(u^c)} s^2_W P_L \bigg)
 u^c_a,
\end{eqnarray}
where $k = 1,2$ are the indices of leptoquark, $T_3 = -1/2$,
$Q_{u^c} = -2/3$ are quark's isospin and electric charge, $Q_S =
-1/3 $ is the electric charge of scalar leptoquarks $S_k$, $a$ and
$i,j$ are quarks and leptons flavor indices and we use $c_{W} =
\cos\theta_W$ and $s_{W} = \sin\theta_W$.
The $\Gamma_{k, S_{L,R}}$ are elements of leptoquark mixing matrix
that brings $S_{L,K}$ to the mass basis $S_k$:
\begin{eqnarray}
S_L = \Gamma^{\dagger}_{S_L, k} S_k, \qquad S^*_R = \Gamma_{k,S_R}
S^*_k,
\end{eqnarray}
where the $S_{L(R)}$ is the field that associates with the
$\overline{e_{j}} P_{L(R)} u^c_a$ terms in
${\mathcal{L}}_{LQ}$~\cite{CHH1999}. Note that in the no-mixing
case ($\Gamma=1$), $S_{1,2}$ reduce to $S_{L,R}$, which are called
chiral leptoquarks, as they only couple to quarks and leptons in
certain chirality structures.
Finally, the couplings $h$ and $h'$ are 3 by 3 matrices, which
give rise to various LFV processes and must be subject to
experimental constraints.

In this work we do not aim at a comprehensive study of the effects
of all possible leptoquark interactions. Instead, we try to
demonstrate that a simple scalar leptoquark model can provide rich
and interesting LFV phenomenons.

\subsection{Muon anomalous magnetic moment $(g-2)_{\mu}$}

\begin{widetext}
\begin{figure}
\begin{center}
\includegraphics[width=12cm]{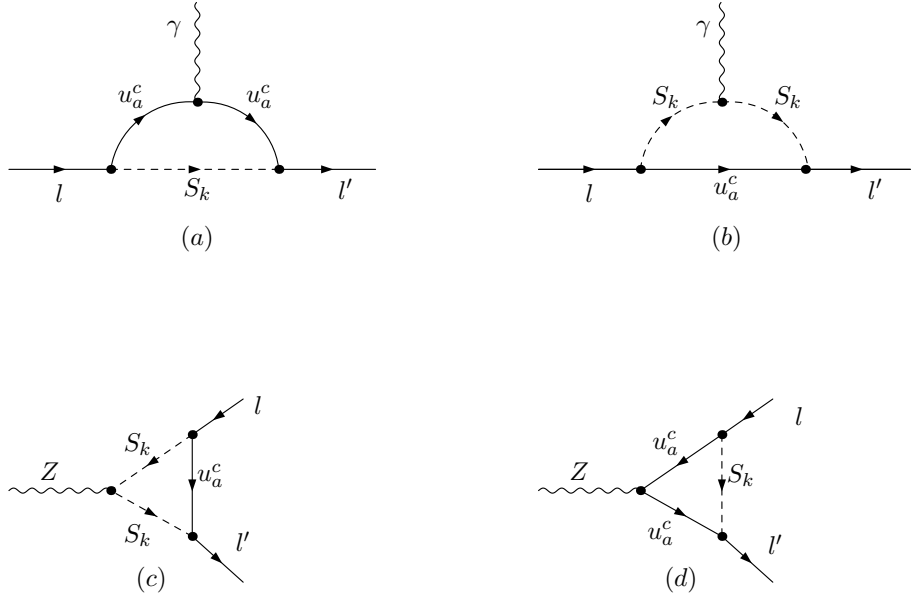}
\end{center}
\caption{Feynman diagrams contributing to $\ell \to \ell' \gamma$
and $Z \to \ell \bar \ell'$,
  ${S}_k$ are the scalar leptoquark $k=1,2$,
  $u^c_a$ are quark up  with  $a=1,2,3$. }
\label{penguin}
\end{figure}
\end{widetext}

The LQ interaction is capable to generate muon anomalous magnetic
moment and resolve the discrepancy between theoretical and
experimental results. The one-loop diagrams are shown in
Fig.~\ref{penguin}(a)-(b) with $l=l'=\mu$. The extra contribution to
$a_\mu $ arises from the LQ model due to quark and scalar
leptoquark one-loop contribution is given by
\begin{widetext}
\begin{eqnarray}
a^{LQ}_{\mu} &=&\no - \frac{N_c m^2_\mu}{8 \pi^2} \sum_{q=1}^{3}
\sum_{k=1}^{2}
 \frac{1}{M^2_{S_k}} \bigg[ \big(|h_{q\mu} \Gamma_{k,S_L}|^2  + |h'_{q\mu} \Gamma_{k,S_R}|^2 \big)
\big( Q_{(u^c)} F_{2}(x)- Q_{S} F_{1}(x) \big) \\&& -
  \frac{m_{(u^c_a)}}{m_\mu}  {\rm Re} \big(h'_{q\mu} h^*_{q\mu} \Gamma^{+}_{S_R,k}\Gamma_{k,S_L} \big)
\big(Q_{(u^c)} F_{3}(x_{ka}) - Q_{S} F_{4}(x_{ka}) \big) \bigg],
\label{eq:a_LQ}
\end{eqnarray}
\end{widetext}
In the above expression, $N_c = 3$, $Q_{S} = -1/3$, $Q_{u^c} =
-2/3$. Our expression agrees with that in \cite{Djouadi:1989md,
Cheung:2001ip}. The kinematic loop functions $F_{i}$ $(i=1,...,4)$
depend on the variable $x = m^2_{(u^c_a)} / m^2_{S_k}$ are given in the appendix A.

Using leptoquark contribution to saturate the deviation given in
Eq.(\ref{amue}), the leptoquark masses $M_{S_{1,2}}$, mixing angle
$\theta_{LQ}$ and couplings $h^{(\prime)}_{q\mu}$, will be
constrained.

\subsection{$\ell \to  \ell' \gamma$}

In this subsection we give the amplitude of $\ell\to \ell'\gamma$ from
leptoquark exchange. According to the gauge invariance, the
amplitude can be written as:
\begin{eqnarray}
i{\mathcal{M}}^{\gamma}&=&ie\bar{u}(p_2) \bigg(
F^{\gamma}_{2RL} P_{L} + F^{\gamma}_{2LR} P_R\bigg) (i\sigma_{\mu\nu}q^\nu)
 u(p_1)\varepsilon^{\mu *}_{\gamma},
\end{eqnarray}
where $\varepsilon_\gamma$ is the polarization vector and $q= p_1
- p_2$ is the momentum transfer. For the amplitude of leptoquark
exchange at one-loop level as depicted in Figure.~\ref{penguin},
we have
\begin{widetext}
\begin{eqnarray}
\label{lepq}
F^{\gamma}_{2LR} &=&\no \frac{N_c}{16 \pi^2}\sum_{q=1}^{3} \sum_{k=1}^{2}
  \frac{1}{M^2_{S_k}}\Bigg[
\big(m_l h'_{q\ell} h^{'*}_{q\ell'} \Gamma^{\dagger}_{S_R,k}
\Gamma_{k,S_R}    + m_{l'} h_{q\ell} h^{*}_{q\ell'}
\Gamma^{\dagger}_{S_L,k} \Gamma_{k,S_L}  \big)
\big(Q_{(u^c)}F_{2}(x)
  - Q_{S}F_{1}(x)\big)
\\ &&- m_{(u^c_a)}
\big(h_{q\ell} h^{'*}_{q\ell'} \Gamma^{\dagger}_{S_R,k}
\Gamma_{k,S_L}\big) \big(Q_{(u^c)}F_{3}(x) -  Q_{S}
  F_4 (x)\big)\Bigg],
  \\
 F^{\gamma}_{2RL} &=& F^{\gamma}_{2LR} ( h
  \leftrightarrow h', R \leftrightarrow L),
\end{eqnarray}
\end{widetext}
with $x = m^2_{(u^c_a)} / m^2_{S_k}$.
The branching ratio of $\ell
\to \ell' \gamma$ is:
\begin{eqnarray}
{\rm Br}(\ell \to \ell' \gamma) &=&\frac{\alpha_{em}}{4
\Gamma(\ell)} \frac{(m^2_\ell - m^2_{\ell'})^3}{ m^3_\ell } \bigg(
|F^{\gamma}_{2LR}|^2 + |F^{\gamma}_{2RL}|^2 \bigg),
\end{eqnarray}
In our numerical calculations we analyze the Brs of the decays
under consideration by using the total decay widths of the
decaying leptons $\Gamma(\ell)$.


\subsection{$Z \to \ell \bar\ell'$ }

The Feynman diagrams of LFV $Z$ decay process are shown in Fig. 1.
The total contribution of all diagrams (c) and (d) can be written
as \beqa \label{ampge} i{\mathcal{M}}^{Z}_{\mu}&=&ie
m^2_Z\bar{u}(p_2) \bigg[\bigg(F^Z_{1L} P_R + F^Z_{1R}P_L\bigg)
\bigg(-g_{\mu\nu} + \frac{q_\mu q_\nu}{m^2_Z} \bigg)\gamma^\nu
\\\no &&
+ \frac{1}{m^2_Z}\bigg(F^Z_{2RL}P_{L} + F^Z_{2LR}P_R\bigg)
(i\sigma_{\mu\nu}q^\nu) \bigg]u(p_1)\varepsilon^{Z}_{\mu} (q)
\eeqa
where $q_\mu$ is the $Z$ four-momentum. The decay rates involve
both $F^Z_{1L(R)}$ and $F^Z_{2LR(RL)}$:
\beqa {\rm Br}(Z \to \ell \bar\ell') &=&
\frac{\alpha_{em}}{6} \frac{m_Z}{\Gamma_Z} \bigg[
\bigg(|F^Z_{1L}|^2 + |F^Z_{1R}|^2 \big) \\\no && + \frac{1}{2
m^2_Z} \bigg(|F_{2LR}(Z)|^2+|F_{2RL}(Z)|^2 \bigg)\bigg],
\eeqa
where the form factors $F^Z_{1L(R)}$ and $F^Z_{2LR(RL)}$ are given
by
\begin{widetext}
\beqa
F^Z_{1L} &=& \frac{N_c}{16 \pi^2} \frac{1}{M^2_{S_k}} \bigg[
h^\prime_{q\ell} h^{\prime *}_{q\ell'} \Gamma^+_{S_R, k}\Gamma_{k, S_R}
\big( g_{S} G_1(x) + g_{R} G_{2}(x) \big)
\\\no &-&\frac{m_{u_a}}{m^2_{\ell} - m^2_{\ell'}}
\big(g_L- g_R \big) \big( h_{q\ell} h^{\prime *}_{q\ell'}
\Gamma^+_{S_R, k}\Gamma_{k, S_L} m_{\ell} - h^\prime_{q\ell}
h^*_{q\ell'} \Gamma^+_{S_L, k}\Gamma_{k, S_R} m_{\ell'}
\big)G_{3}(x)\bigg],
 \\
F^Z_{1R} &=& F^Z_{1L}(h \leftrightarrow h^\prime,  L
\leftrightarrow R ),
 \eeqa
and
\beqa
 F^Z_{2LR} &=& \frac{N_c}{16 \pi^2}\frac{1}{M^2_{S_k}} \bigg[
h_{q\ell} h^{\prime *}_{q\ell'} \Gamma^{\dagger}_{S_R,
k}\Gamma_{k,S_L} m_{u_a}(g_R + g_L)G_3(x)
 \\\no
&+& \big( g_R
 h^\prime_{q\ell} h^{\prime *}_{q\ell'} \Gamma^{\dagger}_{S_R, k}\Gamma_{k,S_R}
m_{\ell} + g_L h_{q\ell} h^{ *}_{q\ell'} \Gamma^{\dagger}_{S_L,
k}\Gamma_{k,S_L} m_{\ell'} \big)G_4(x)
 \\\no
 &-& g_{S}\bigg( (
h^\prime_{q\ell} h^{\prime *}_{q\ell'} \Gamma^{\dagger}_{S_R,
k}\Gamma_{k,S_R} m_{\ell} + h_{q\ell} h^{ *}_{q\ell'}
\Gamma^{\dagger}_{S_L, k}\Gamma_{k,S_L} m_{\ell'}) G_5(x)
+ m_{u_a} h_{q\ell} h^{ \prime *}_{q\ell'} \Gamma^{\dagger}_{S_R,
k}\Gamma_{k,S_L} G_6(x)\bigg)\bigg],
 \\
 F^Z_{2RL} &=& F^Z_{2LR}(h \leftrightarrow h^\prime ,L
 \leftrightarrow R ),
 \eeqa
\end{widetext}
where we have $x = m^2_{u_a} / m^2_{S_k}$ and the couplings
$g_{R,L}$ and $g_S$ are given by
\begin{eqnarray}
g_R &=& -\frac{2}{\sin\theta_W \cos\theta_W} \big ( T_{3(u^c)} - Q_{(u^c)} \sin^2\theta_W \big), \\
g_L &=&  Q_{(u^c)} \tan\theta_W, \qquad g_S =  Q_{S} \tan\theta_W.
\end{eqnarray}
In the above expressions of $F^Z_{1L(R)}$, we keep only the leading term
in $m_Z^2/m_{S_k}^2$. The explicit expressions of one loop
functions $G_n$ $(n=1,..6)$ can be found in the appendix A.

\section{Numerical results and discussion}

We are now ready to give some numerical results. The quark mass
are evaluated  at the scale of the $\mu =
300$~GeV~\cite{Gray:1990}, which is the typical leptoquark mass
used in this work,
\begin{eqnarray}
m_t = 161.4 \,{\rm GeV}, \quad m_c = 0.55\, {\rm GeV},\quad
 m_u  = 11.4 \times 10^{-3}\,{\rm GeV},
\end{eqnarray}
and for the following quantities we use~\cite{pdg}
\begin{eqnarray}
\alpha_{em} = 1/137.0359, \quad M_W = 80.45 \,{\rm GeV} , \quad M_Z = 91.1875 \,{\rm GeV}.
\end{eqnarray}
For simplicity, we assume that the couplings $h$ and $h'$ are real
and equal to each other, i.e.
\begin{eqnarray}
h = h^\prime = h^*.
\end{eqnarray}
We use leptoquark mass splitting $\Delta= 500$ GeV in our
analysis, where $\Delta$ is defined as $\sqrt{M^2_{S_2} -
M^2_{S_1}}$. Consequently, the remaining parameters in the
leptoquark model are the mass of the light scalar leptoquark
$M_{S_1}$, the mixing angle $\theta_{LQ}$, and the couplings
$h_{q\ell}$.

\subsection{Muon Anomalous Magnetic Moment $a_\mu$}

\begin{figure*}[ht]
  \begin{tabular}{cc}
    \resizebox{90mm}{!}{{\hspace{-3cm}}\includegraphics{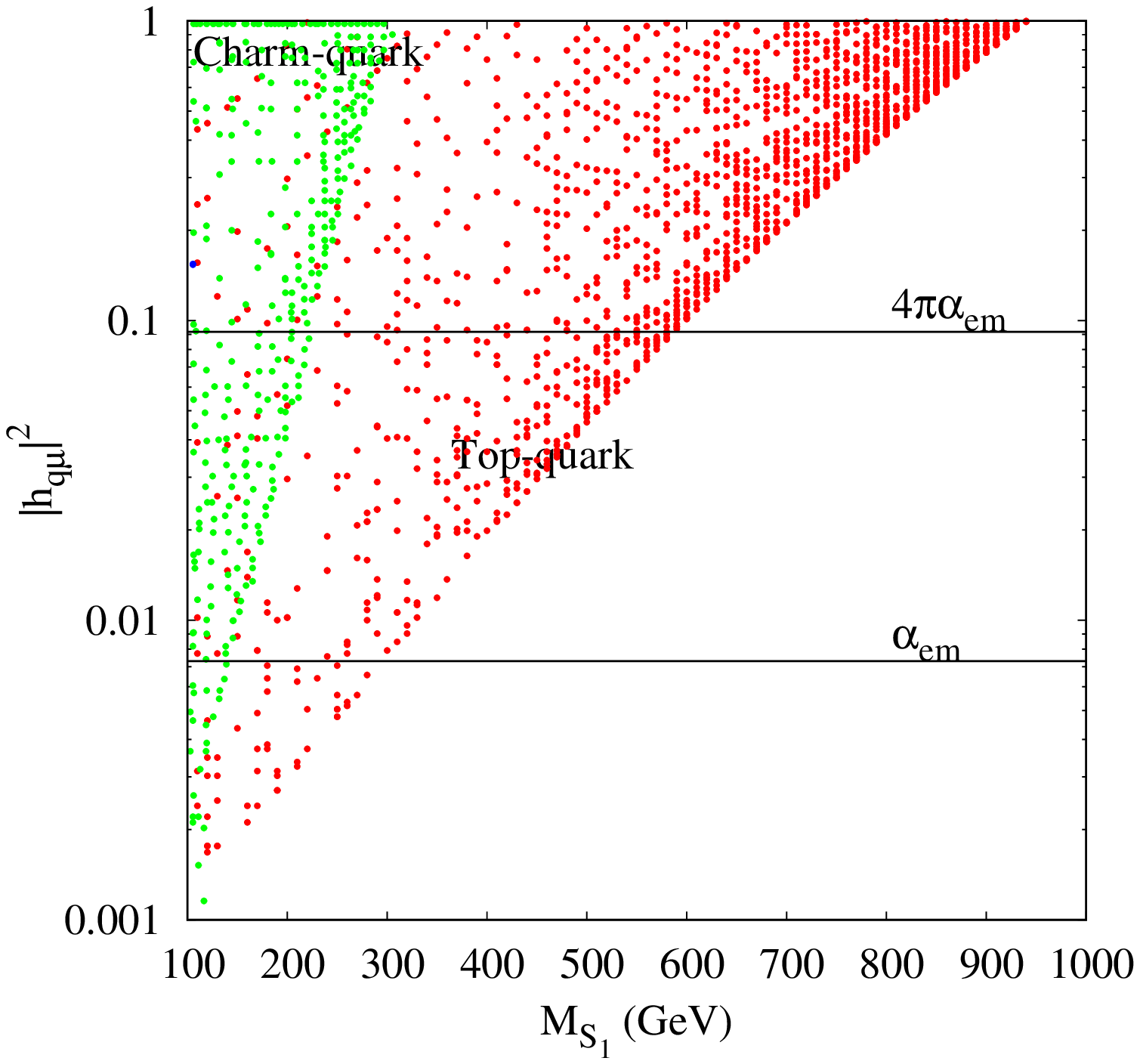}}&{\hspace{-0.5cm}}
    \resizebox{90mm}{!}{{\hspace{-3cm}}\includegraphics{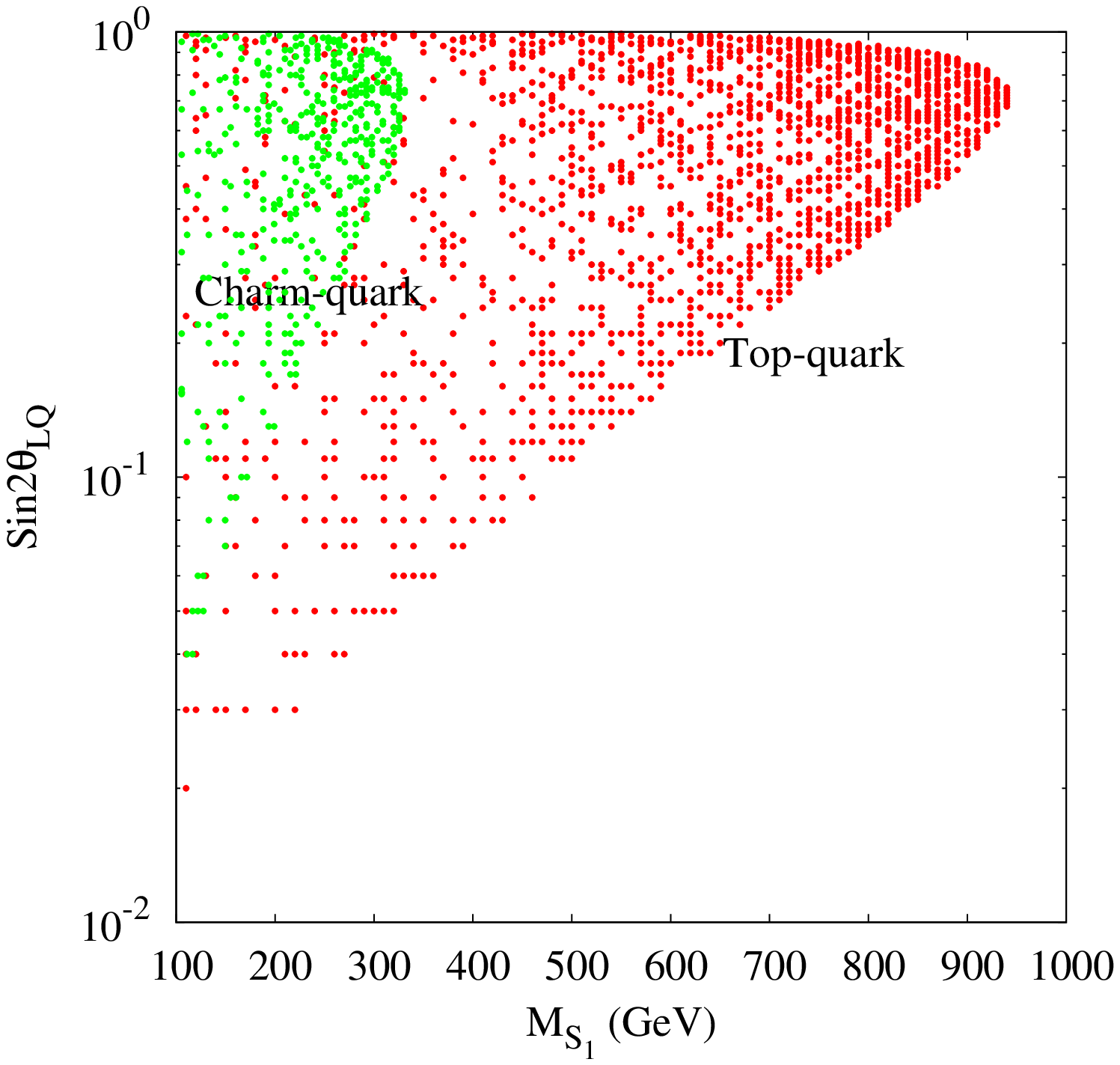}}
\end{tabular}
\caption{ Scatter plot in the plane ($M_{S_1}- |h_{q\mu}|^2$) in
 the left panel,($M_{S_1}- \sin2\theta_{LQ}$) in the right panel. These are
 allowed regions in the parameter space that give $a^{LQ}_\mu=\Delta
 a_\mu=(295\pm87.7)\times10^{-11}$.}
 \label{figure1}
\end{figure*}

In this section we discuss a few phenomenological aspects of the
leptoquark contributions to $a_\mu$.
In the left panel of Fig.~\ref{figure1}, we present an scatter
plot in the $(M_{S_1}- |h_{q\mu}|^2)$ plane for top quark
contribution (red) and charm quark contribution (green), which are
allowed by $a^{LQ}_\mu=\Delta a_\mu=(295\pm87.7)\times10^{-11}$
[see Eq. (\ref{amue})] within the 1$\sigma$ range of data. We note
that it is not possible to use the up quark loop contribution
alone for the $a_\mu^{LQ}=\Delta a_\mu$, since the mixing angle
and couplings $h_{u\mu}$ are strongly constrained by the $\pi$
leptonic decays (see Appendix B).

In order to see the impact of the mixing angle, we present in the
right panel of Fig.\ref{figure1} the allowed regions
$a^{LQ}_{\mu}=\Delta a_\mu$ in the $(M_{S_1}- \sin2\theta_{LQ})$
plane. We use $\alpha_{em}\leq h^2_{q\mu}\leq1$. The contribution
dominates around $\sin2\theta_{LQ} \sim 0.7$ both for top and
charm quark contributions. We see that the constraint from
$a_{\mu}$ confines the allowed range of $M_{S_1}$ to $M_{S_1}
\lesssim 950 $ GeV for top quark contribution and to $M_{S_1}
\lesssim 350 $ GeV for charm quark contribution at the 1$\sigma$
level. These parameter space will be used for later study of LFV
processes. The light leptoquark mass should be below 1 TeV, if
leptoquarks with couplings of electromagnetic strength are
responsible to the deviation $\Delta a_\mu$. It is interesting
that LHC may have good chance to observe these
particles~\cite{LHC}.

\subsection{Lepton Flavor Violating $l\to l'\gamma$ and $Z\to l\bar l'$ Decays}
In this section, we investigate the LFV decay processes generated
by the same leptoquark scalar interactions. We consider only
parameter space that corresponds to $a^{LQ}_\mu=\Delta a_\mu$ when
it is appropriate. We discuss $\mu \to e \gamma$ and $\tau \to e
\gamma,\,\mu\gamma$ decays first.

\begin{center}
\begin{figure*}
  \begin{tabular}{cc}
    \resizebox{90mm}{!}{{\hspace{-3cm}}\includegraphics{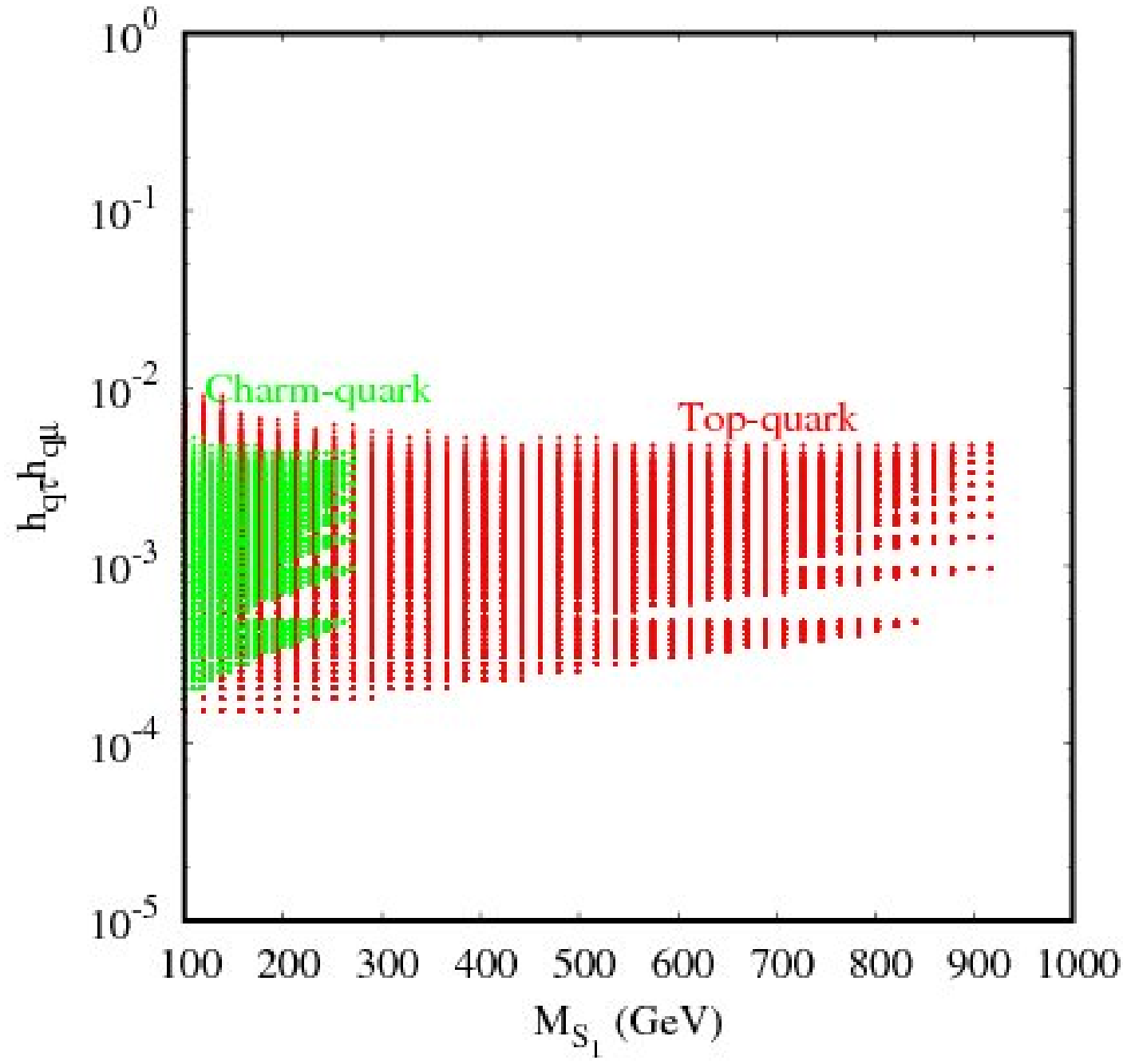}} &{\hspace{-0.5cm}}
    \resizebox{90mm}{!}{{\hspace{-3cm}}\includegraphics{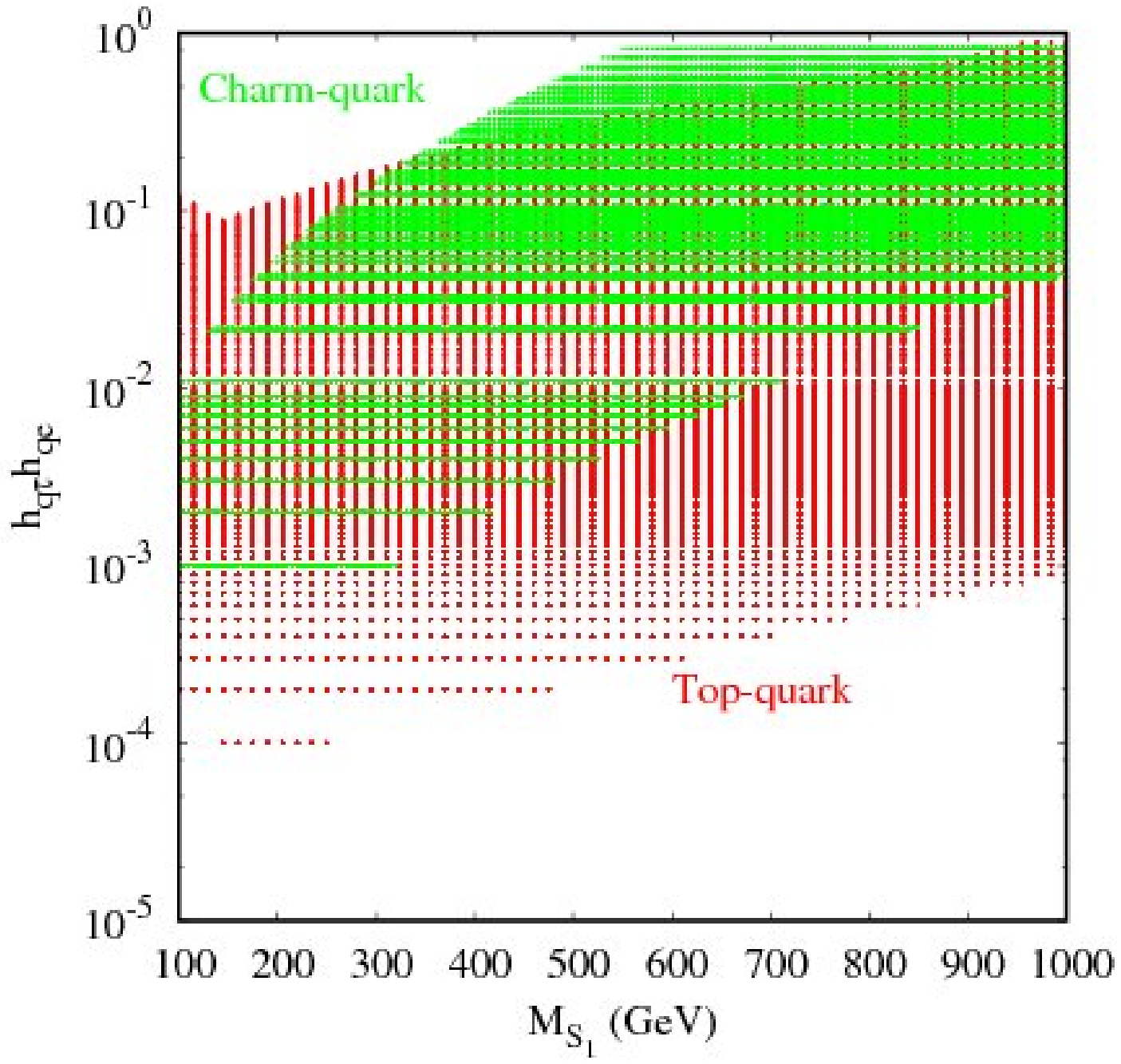}}
  \end{tabular}
\caption{
Scatter plots of leptoquark parameters in
 $(M_{S_1}, h_{q\ell}h_{q\ell'})$ planes from $(\ell \to
\ell' \gamma)$ bounds given in Eq.~(\ref{eq:ll'g}). The left
(right) figure is for the
 $\tau \to \mu \gamma$ ($\tau \to e \gamma$) case with top and charm quark contributions.
} \label{figure3}
\end{figure*}
\end{center}

\begin{center}
\begin{figure*}
\includegraphics[width=11cm]{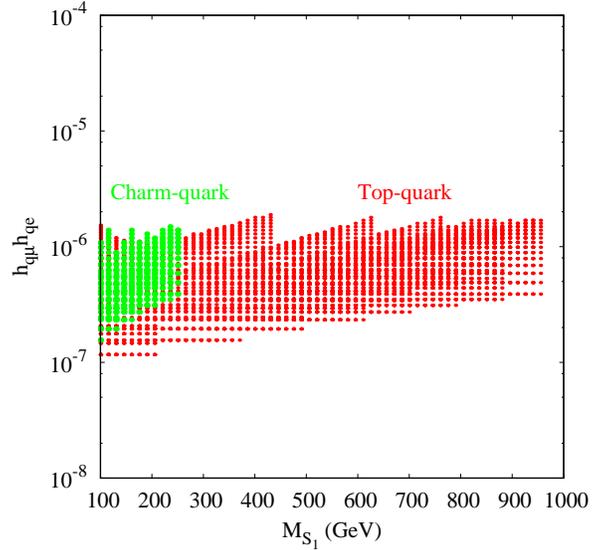}
\caption{Same as Fig.~\ref{figure3} except for the $\mu \to e
\gamma$ case.} \label{figure4}
\end{figure*}
\end{center}

In Fig.~\ref{figure3} and Fig.~\ref{figure4} we show scatter plots
of the allowed parameters in $(M_{S_1},
h_{q\ell}h_{q\ell^\prime})$ planes from bounds of $\tau \to
\mu\gamma$, $\tau \to e \gamma$ and $\mu \to e \gamma$ rates. Note
that in the plots we use
 \beqa
 1.5\times 10^{-13}&\leq&{\rm Br}(\mu \to e \gamma) < 1.2 \times 10^{-11},
 \no\\
 1\times 10^{-9}&\leq&{\rm Br}(\tau \to e \gamma) < 1.1 \times 10^{-7},
 \no\\
 1\times 10^{-9}&\leq&{\rm Br}(\tau \to \mu \gamma) < 6.8 \times 10^{-8}.
 \label{eq:ll'g}
 \eeqa
where the upper bounds are from the current limits:
Eqs.~(\ref{lfv1})-(\ref{lfv3}), while the lower bound for ${\rm
Br}(\mu \to e \gamma)$ is from \cite{LQll'gamma} and the lower
bounds for $\tau\to l\gamma$ are for illustration. For the
$\tau\to \mu\gamma$ and $\mu\to e\gamma$ cases the $(g-2)_\mu$
constraint is taken into account.

For different quark contribution the couplings are bounded in the
following ranges: $10^{-4}\lesssim h_{q\tau}h_{q\mu}\lesssim
10^{-2}$, $10^{-3}\lesssim h_{c\tau}h_{ce}\lesssim 1$,
$10^{-4}\lesssim h_{t\tau}h_{te}\lesssim 1$ and $10^{-7}\lesssim
h_{q\mu}h_{qe}\lesssim 10^{-6}$. For the $\tau\to\mu \gamma$ and
$\mu\to e\gamma$ cases, the allowed leptoquark masses are
$m_{S_1}\lesssim 250-300$ GeV and 1~TeV for $c$-quark and $t-$
quark loop contribution, respectively. These region are determined
from the bounds and the muon $g-2$ constraint (see also
Fig.~\ref{figure1}) at the same time. On the other hand the
couplings governing $\tau\to e\gamma$ decay and those generating
muon $g-2$ contribution are decoupled, the parameters
corresponding to the former bounds are free from the latter
constraint. The resulting allowed regions are larger in this
cases.
The parameters in these allowed regions will be used to predict
$Z\to l\bar l'$ decays. To have an idea of the size the allowed
couplings, we give that upper bound on $h_{q\ell} h_{q\ell'}$
obtained form the present $l\to l'\gamma$ limits in
Table~\ref{TBC2}. We see that the $\mu\to e\gamma$ constraint is
more effective in restricting the sizes of $h_{q\ell} h_{q\ell'}$.

\begin{table}[h]
\begin{center}
\begin{tabular}{|c|c|c|}
\hline\hline
Decay mode & $h_{c\ell}h_{c\ell'}$ & $h_{t\ell}h_{t\ell'}$ \\
\hline\hline
$ \tau \to \mu \gamma $  & $ < 5.29 \times 10^{-3}$ & $ < 9.11 \times 10^{-3}$ \\
$ \tau \to e \gamma $  & $ < 0.81 $ & $ < 0.82$ \\
$ \mu \to e \gamma $  & $< 1.45 \times 10^{-6}$ & $ < 1.92 \times 10^{-6}$  \\
\hline\hline
\end{tabular}
\caption{ Constraints on the parameters $h_{q\ell}h_{q\ell'}$
$(q=t,c)$ coming from radiative FCNC processes induced by the
scalar leptoquark using the present experimental bounds.}
\label{TBC2}
\end{center}
\end{table}

\begin{center}
\begin{figure*}
  \begin{tabular}{cc}
    \resizebox{90mm}{!}{{\hspace{-3cm}}\includegraphics{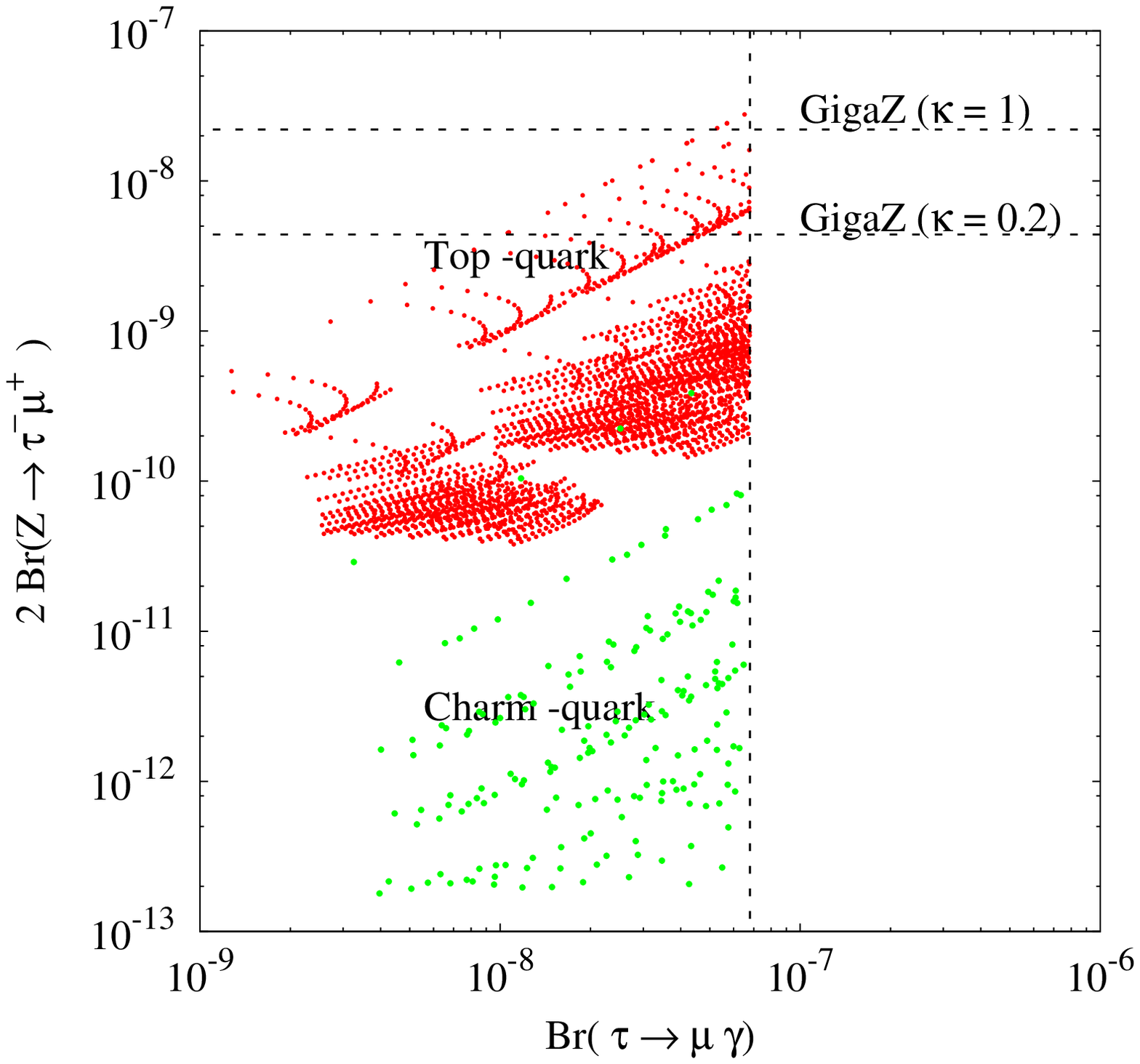}} &{\hspace{-1cm}}
    \resizebox{90mm}{!}{{\hspace{-3cm}}\includegraphics{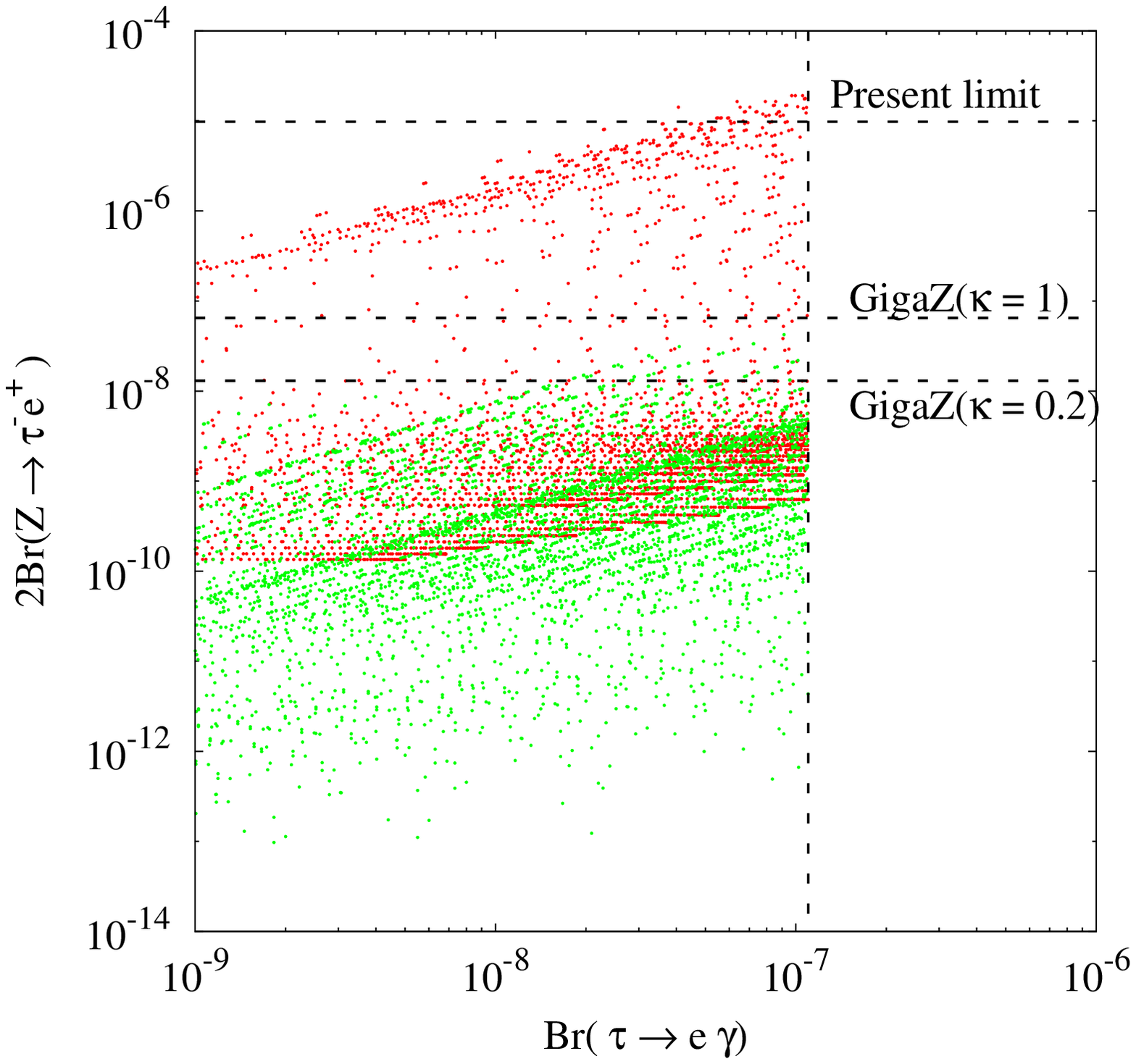}}
  \end{tabular}
\caption{ The correlation between Br$(\tau \to \ell' \gamma)$ and
Br$(Z \to \tau \ell')$ where $\tau = e , \mu$.} \label{figure7}
\end{figure*}
\end{center}

\begin{center}
\begin{figure*}
\includegraphics[width=11cm]{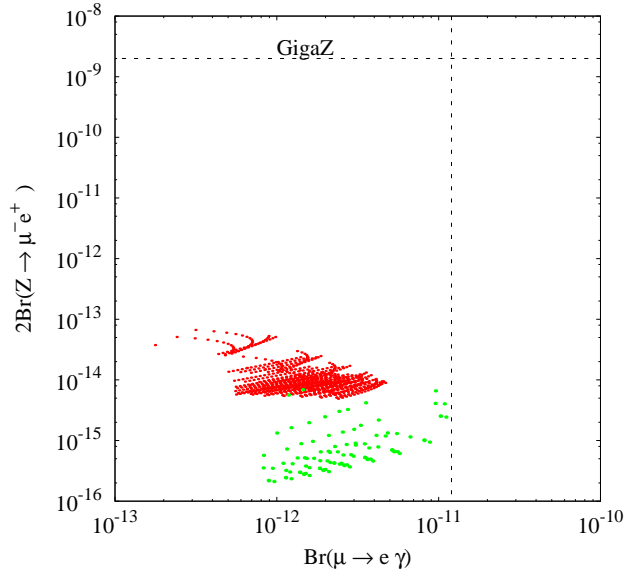}
\caption{ The correlation between Br$(\mu \to e \gamma)$ and Br$(Z
\to \mu e)$.} \label{figure8}
\end{figure*}
\end{center}

In Fig.~\ref{figure7} and \ref{figure8}, we give the predicted
$Z\to l\bar l'$ rates in correlation with Br$(l \to l' \gamma)$.
We see that $Br(Z \to \tau^\mp e^\pm)$ can reach $1.95 \times
10^{-5}$, which is comparable with the present bound, and $Br(Z
\to \mu^\mp \tau^\pm)$ can reach $2.34 \times 10^{-8}$, which will
be accessible by future linear colliders. On the contrary, the
current bound on the $\mu\to e\gamma$ decay imposes very strong
constraints on the related couplings as shown in Table I. Hence
the predicted $Br(Z \to \mu^\mp e^\pm)$ is rather small and is too
low to be observed in the near future.
In Fig.~\ref{figure7}, we see that the $Z\to l\bar l'$ rates are
roughly positively correlating with the $l \to l' \gamma$ rates
and the top quark loop contributions are larger than the charm
quark's ones. To have observable $Z\to \tau^\mp\mu^\pm$ and $Z\to
\tau^\mp e^\pm$, the $\tau \to\mu\gamma,\,e\gamma$ rates are
predicted to be close to the present bounds.

In this work the analysis has been performed for the scalar
leptoquark case. It is possible that vector leptoquarks may also
contribute to $(g-2)_\mu$ and LFV processes. As shown in
Ref.~\cite{Bigi:1985jq,DBC}, quite often $(g-2)_\mu$ and LFV
processes provide more stringent constraints on vector leptoquark
couplings and masses than on scalar leptoquark ones. For example,
using the measured $m_t$ and the formula given in
\cite{Bigi:1985jq}, the present $\Delta a_\mu$ leads to a very
large mass scale $\Lambda\simeq 500$~TeV in the vector leptoquark
case, where $\Lambda$ was defined from the relation:
$4\pi/\Lambda^2\equiv g^2_{LQ}/m^2_{LQ}$. The mass scale is much
larger than the corresponding mass scale exhibited in Fig.~2,
which is found to be $\Lambda\simeq$ few -- ${\mathcal
O}(10)$~TeV. Similarly, in $l\to l'\gamma$ processes, the
constraints on vector leptoquark parameters are usually more
severe~\cite{DBC}.


\section{Conclusion}

Motivated by the reported discrepancy of the muon $g-2$ results,
we studied the lepton flavor violating $\ell \to \ell' \gamma$ and
$Z \to \ell \bar\ell'$ decays in the LQ model. We showed that the
$g-2$ anomaly favors LQ masses in rather low-energy regime, e.g.
$< 1$ TeV, which is within the reach of the forthcoming Large
Hadron Collider.

We found that leptoquarks can generate sizable LFV $l\to l'\gamma$
decays. The present experimental limits are used to confined the
leptoquark parameter space. On the other hand, it is interesting
to search for these LFV effects in experiments, such as MEG, B
factories and the super B factory.

We predict $Br(Z \to \tau^\mp e^\pm)$ reaching $10^{-5}$ and $Br(Z
\to \mu^\mp \tau^\pm)$ reaching $2\times 10^{-8}$, which can be
accessible by present experiments and future linear colliders,
such as ILC. On the contrary, the current bounds on LFV impose
very strong constraints on the $Br(Z \to \mu^\mp e^\pm)$ and the
ratio is too low to be observed in the near future. In this case,
it is much useful to search for the LFV effects in $\mu\to
e\gamma$ decay.

\section{Acknowledgment}
The authors would like to thank Dr. Stefan Ritt for useful
discussions. This work is supported in part by the National
Science Council of R.O.C under grant number: NSC96-2811-M-033-005
and NSC-95-2112-M-033-MY2.


\appendix

\section{One loop functions}
The loop functions $F_i$ and $G_i$ used in Sec. II are given by
\begin{eqnarray}
F_{1}(x) &=& \frac{\big[2 + 3x -6x^2+x^3+6x\log(x)\big]}{12(1-x)^4},\\
F_{2}(x) &=& \frac{\big[1 - 6x +3x^2+2x^3-6x^2\log(x)\big]}{12(1-x)^4},\\
F_{3}(x) &=& \frac{-1}{2(1-x)^3}\big[ 3 -4x + x^2 + 2 \log(x)\big],\\
F_{4}(x) &=& \frac{1}{2(1-x)^3}\big[ 1 - x^2 + 2 x \log(x)\big],
\end{eqnarray}
and
\begin{eqnarray}
G_1(x) &=& \frac{\big[ -2 + 9 x^2 -18 x^4 + 11 x^6 - 12 x^6 \log(x) \big]}{36(x^2-1)^4} ,\\
G_2(x) &=& \frac{1}{36(x^2-1)^4}\\ \no & \times &
\big[ 16 - 45 x^2 + 36 x^4 - 7 x^6 + 12 (-2+3 x^2) \log(x)\big]\\
G_3(x) &=& \frac{ 3 - 4 x^2 + x^4 + 4 \log(x) }{4(x^2-1)^3} ,\\
G_4(x) &=& \frac{ 2 + 9 x^2 -6 x^4 + x^6 + 12 x^2 \log(x) }{12(x^2-1)^4} ,\\
G_5(x) &=& \frac{1 - 6 x^2 + 3 x^4 + 2 x^6 - 12 x^4 \log(x) }{12(x^2-1)^4} ,\\
G_6(x) &=& \frac{1}{2(x^2-1)^3} \big[ -1 + x^4 - 4 x^2\log(x)
\big].
\end{eqnarray}

\section{Constraint form $\pi \to e\nu_{e}$ and $\pi \to \mu \nu_{\mu}$ decays }

We follow \cite{pi_decay,DBC} to constrain leptoquark parameters
using pion decay data. Form the interactions given in
Eq.~(\ref{lag}), we obtain the effective four-Fermi interaction
\begin{eqnarray}
{\mathcal{L}}_{eff} &=& \label{lag2}
-\frac{h'_{ai}h^{'*}_{bj}\Gamma^+_{R,k} \Gamma_{k,R}}{M^2_{S_k}}
(\bar{e}^c_iP_L u_a )(\bar{d}_b P_R \nu^c_j) \\\no &-&
\frac{h_{ai} h^{'*}_{bj}\Gamma^\dagger_{R,k}
\Gamma_{k,L}}{M^2_{S_k}} (\bar{e}^c_iP_R u_a )(\bar{d}_b P_R
\nu^c_j)
\end{eqnarray}
By using the Fierz transformation, we can rewite the Eq.(\ref{lag2}) as
\begin{eqnarray}
{\mathcal{L}}_{eff} &=&\no  -\frac{1}{2 M^2_{S_k}}
h'_{ai}h^{'*}_{bj} \Gamma^\dagger_{R,k} \Gamma_{k,R}(\bar{d}_{L,b}
\gamma_\mu u_{L,a})(\bar{\nu}_{L,j} \gamma^\mu e_{L,i})
\\ &+&\frac{1}{2M^2_{S_k} }h_{ai}h^{'*}_{bj}\Gamma^\dagger_{R,k}
\Gamma_{k,L} (\bar{d}_{L,b}  u_{R,a})(\bar{\nu}_{L,j} e_{R,i})
\end{eqnarray}
On the other hand, the conventional interation for the $\pi \to l\nu_{l}$ decay in the SM is given by
\begin{eqnarray*}
{\mathcal{L}}_{eff} &=& -\frac{G_F V_{ud}}{\sqrt{2}}[\bar{\nu}\gamma_\mu (1-\gamma_5)l]
[\bar{d}\gamma^\mu (1-\gamma_5)u] + {\rm h.c}
\end{eqnarray*}
here $|V_{ud}|$ is the Cabibbo-Kobayashi-Maskawa (CKM) matrix elements between the constituent of the pion meson, $G_F$ is the Fermi couplings constant. The ratio $R_{th}$ of the electronic and muonic decay modes is \cite{Cirigliano:2007xi}
\begin{eqnarray}
R_{th} &=& \no \frac{\Gamma_{SM}(\pi^+ \to \bar{e}\nu_e)}{\Gamma_{SM}(\pi^+ \to \bar{\mu}\nu_\mu)}
\\\no
&=& \bigg(\frac{m^2_e}{m^2_\mu}\bigg)\bigg(\frac{m^2_\pi - m^2_e}{m^2_\pi - m^2_\mu}\bigg)^2\bigg(1 + \delta\bigg)
\\
&=& (1.2352 \pm 0.0001) \times 10^{-4}
\end{eqnarray}
where $\delta$ is the radiative corrections, Thus the ratio
$R_{th}$ is very sensitive to non standard model effects (such as
multi-Higges, non-chiral leptoquarks).
The experimental ratio is \cite{pdg}
\begin{eqnarray}
R_{exp} = (1.2302 \pm 0.004) \times 10^{-4}
\end{eqnarray}
The interference between the standard model and LQ model can be expressed by
\begin{widetext}
\begin{eqnarray}
R_{SM-LQ} &=& R_{th} +R_{th}\,\frac{m^2_{\pi^+}}{m_u + m_d}
\bigg(\frac{1}{\sqrt{2}}\frac{{\rm Re} (h_{ue}h^{'*}_{ue})}{G_F
V_{ud}M^2_{S_k}} \frac{1}{m_{e}}-\frac{1}{\sqrt{2}} \frac{{\rm Re}
(h_{u\mu}h^{'*}_{u\mu})}{G_F
V_{ud}M^2_{S_k}}\frac{1}{m_{\mu}}\bigg)
\Gamma^\dagger_{R,k}\Gamma_{k,L}
\end{eqnarray}
At 2$\sigma$ level, we get
\begin{eqnarray}
 R_{min} < \sum^{2}_{k=1}\bigg(\frac{m_{\pi}}{m_{e}} \frac{{\rm Re}
 (h_{ue}h^{'*}_{ue})}{M^2_{S_k}}-\frac{m_{\pi}}{m_{\mu}} \frac{{\rm Re} (h_{u\mu}h^{'*}_{u\mu})}{ M^2_{S_k}}\bigg)
 \Gamma^\dagger_{R,k}\Gamma_{k,L} < R_{max}
\end{eqnarray}
\end{widetext}
where,
\begin{eqnarray}
R_{min} &=& -1.06 \times 10^{-8} {\rm GeV}^{-2},\\  R_{max}  &=& 2.45 \times 10^{-9} {\rm GeV}^{-2}.
\end{eqnarray}
The total contribution to $R_{SM-LQ}$ must be smaller than the differences
between SM and experiment within the error limits allow.
\newpage



\begin{thebibliography}{99}


\bibitem{Bennet}
  G.~W.~Bennett {\it et al.}  [Muon g-2 Collaboration],
  Phys.\ Rev.\ Lett.\  {\bf 89}, 101804 (2002)
  [Erratum-ibid.\  {\bf 89}, 129903 (2002)]
  [arXiv:hep-ex/0208001];
  Phys.\ Rev.\ Lett.\  {\bf 92}, 161802 (2004)
  [arXiv:hep-ex/0401008].



\bibitem{Miller}
  J.~P.~Miller, E.~de Rafael and B.~L.~Roberts,
  Rept.\ Prog.\ Phys.\  {\bf 70}, 795 (2007)
  [arXiv:hep-ph/0703049].


\bibitem{Aoyama:2007mn}
  T.~Aoyama, M.~Hayakawa, T.~Kinoshita and M.~Nio,
  Phys.\ Rev.\  D {\bf 77}, 053012 (2008); T.~Kinoshita and M. Nio, Phys.\ Rev. {\bf 73}, 013003 (2006);
  T.~Kinoshita and M. Nio, Phys.\ Rev. {\bf 70}, 113003 (2004).



\bibitem{Bigi:1985jq}
I.~I.~Y.~Bigi, G.~Kopp and P.~M.~Zerwas,
  Phys.\ Lett.\  B {\bf 166}, 238 (1986).





\bibitem{Mahanta:2001yc}
  U.~Mahanta,
  Eur.\ Phys.\ J.\  C {\bf 21}, 171 (2001)
  [Phys.\ Lett.\  B {\bf 515}, 111 (2001)]
  [arXiv:hep-ph/0102176].

\bibitem{Cheung:2001ip}
  K.~m.~Cheung,
  Phys.\ Rev.\  D {\bf 64}, 033001 (2001)
  [arXiv:hep-ph/0102238].


\bibitem{ps}
  J.~C.~Pati and A.~Salam,
  Phys.\ Rev.\  D {\bf 10}, 275 (1974)
  [Erratum-ibid.\  D {\bf 11}, 703 (1975)].




\bibitem{GUT}
  H.~Georgi and S.~L.~Glashow,
  Phys.\ Rev.\ Lett.\  {\bf 32}, 438 (1974);
  H. Georgi, AIP Conf. Proc. {\bf 23}, 575 (1975);
  H.~Fritzsch and P.~Minkowski,
  Annals Phys.\  {\bf 93}, 193 (1975).




\bibitem{lowLQ}
  E.~Farhi and L.~Susskind,
  Phys.\ Rept.\  {\bf 74}, 277 (1981);
  K.~D.~Lane and M.~V.~Ramana,
  Phys.\ Rev.\  D {\bf 44}, 2678 (1991);
  B.~Schrempp and F.~Schrempp,
  Phys.\ Lett.\  B {\bf 153}, 101 (1985).


\bibitem{Lagr1}
  W.~Buchmuller, R.~Ruckl and D.~Wyler,
  Phys.\ Lett.\  B {\bf 191}, 442 (1987)
  [Erratum-ibid.\  B {\bf 448}, 320 (1999)].


\bibitem{pi_decay}
  O.~U.~Shanker,
  Nucl.\ Phys.\  B {\bf 204}, 375 (1982).


\bibitem{pp}
  O.~J.~P.~Eboli and A.~V.~Olinto,
  Phys.\ Rev.\  D {\bf 38}, 3461 (1988);
  J.~L.~Hewett and S.~Pakvasa,
  {\it ibid.}  D {\bf 37}, 3165 (1988);
  J.~Ohnemus, S.~Rudaz, T.~F.~Walsh and P.~M.~Zerwas,
  Phys.\ Lett.\  B {\bf 334}, 203 (1994)
  [arXiv:hep-ph/9406235].



\bibitem{ep}
 J. Wudka, Phys.Lett., B{\bf 167}, 337 (1986); M.A. Doncheski and
 J.L. Hewett, Z.Phys. C{\bf 56}, 209 (1992).

\bibitem{e+e-}
 J.L Hewett and T.G. Rizzo, Phys.Rev. D{\bf 36}, 3367 (1987);
 J.L Hewett and S. Pakvasa, Phys.Lett. B{\bf 227}, 178 (1987);
 J.E. Cieza and O.J.P. \'Eboli, Phys.Rev. D{\bf 47}, 837 (1993).

\bibitem{Djouadi:1989md}
  A.~Djouadi, T.~Kohler, M.~Spira and J.~Tutas,
  Z.\ Phys.\  C {\bf 46}, 679 (1990).

\bibitem{Lagr2}
 J. Bl\"umlein and R. R\"uckl, Phys.Lett. B{\bf 304}, 337 (1993).

\bibitem{e-gamma}
 O.J.P. \'Eboli et al., Phys.Lett. B{\bf 311}, 147 (1993);
 H. Nadeau and D. London, Phys.Rev. D{\bf 47}, 3742 (1993).

\bibitem{pdg}
 W. M. Yao et {\it al} Particle Data Group, { J. Phys. G}{\bf 30},
 1 (2006) and 2007 partial update for the 2008 edition.


\bibitem{gamma-gamma}
 G. B\'elanger, D. London and H. Nadeau, Phys.Rev. D{\bf 49}, 3140 (1993).

\bibitem{DBC}
 S. Davidson, D. Bailey and A. Campbell,
 Z.Phys. C{\bf 61}, 613 (1994).

\bibitem{Leurer}
 M. Leurer, Phys.Rev.Lett.{\bf 71} (1993) 1324;
 Phys.Rev. D{\bf 49}, 333 (1994);
 Phys.Rev. D{\bf 50}, 536 (1994).

\bibitem{bw86}
 W. Buchm\"uller and D. Wyler, Phys.Lett. B{\bf 177}, 377 (1986).

\bibitem{CHH1998}
  C.~K.~Chua and W.~Y.~Hwang,
  Phys.\ Rev.\  D {\bf 60}, 073002 (1999)
  [arXiv:hep-ph/9811232].

\bibitem{CHH1999}
 C.~K.~Chua, X.~G.~He and W.~Y.~Hwang,
  Phys.\ Lett.\  B {\bf 479}, 224 (2000)
  [arXiv:hep-ph/9905340].

\bibitem{Benbrik:2008ik}
  R.~Benbrik and C.~H.~Chen,
  arXiv:0807.2373 [hep-ph].

\bibitem{Gray:1990}
N.~Gray, D.~J.~Broadhurst, W.~Grafe and K.~Schilcher, Z.~Phys.
C{\bf 48}(1990) 673;
%
K.~G.~Chetyrkin and M.~Steinhauser, Phys. Rev. Lett {\bf 83}
(1999) 4001;
%
{\it ibid.} Nucl. Phys. B{\bf 573} (2000) 617; S.~G.~Gorishny,
A.~L.~Kataev, S~.A~.Larin and L.~R.~Surguladze, Mod. Phys. Lett.
A{\bf 5}(1990)2703,{\it ibid.} Phys. Rev. D{\bf 43} (1991) 1633.


\bibitem{LHC}
 V.~A.~Mitsou, N.~C.~Benekos, I.~Panagoulias and T.~D.~Papadopoulou,
  Czech.\ J.\ Phys.\  {\bf 55}, B659 (2005)
  [arXiv:hep-ph/0411189];
 S.~Abdullin and F.~Charles,
  Phys.\ Lett.\  B {\bf 464}, 223 (1999)
  [arXiv:hep-ph/9905396].

\bibitem{LQreview}
 S. Rolli and M. Tanabashi, in \cite{pdg}.

\bibitem{LQll'gamma}
 S.~Ritt  [MEG Collaboration],
  Nucl.\ Phys.\ Proc.\ Suppl.\  {\bf 162} (2006) 279;
 http://meg.web.psi.ch/.


\bibitem{Riemann}
 T. Riemann and G. Mann,
 in {\it Proc. of the Int. Conf. Neutrino'82, 14-19 June
 1982, Balatonf{\"u}red, Hungary} (A. Frenkel and E. Jenik, eds.),
 vol. II, pp. 58-, Budapest, 1982,
 scanned copy at
 http://www.ifh.de/$\sim$riemann, G. Mann and T. Riemann, {\it
 Annalen Phys.} {\bf 40}, 334 (1984).

\bibitem{Ganapathi}
 V. Ganapathi, T. Weiler, E. Laermann, I. Schmitt, and P. Zerwas, {\it Phys.  Rev.} {\bf D27} (1983) 579;
 M. Clements, C. Footman, A. Kronfeld, S. Narasimhan, and D. Photiadis, {\it Phys.
 Rev.} {\bf D27} (1983) 570;
 M. A .Perez, G. T. Velasco, J. J. Toscano, {\it Int.J.Mod.Phys.} {\bf A19} (2004) 159;
 A. Flores-Tlalpa, J.M. Hernandez, G. Tavares-Velasco, J.J. Toscano, {\it Phys. Rev.} {\bf
 D65}, 073010 (2002).

\bibitem{Illana}
 J. I. Illana, M. Jack and T. Riemann, hep-ph/0001273, (2000);
 J. I. Illana, and T. Riemann, {\it Phys. Rev.} {\bf D63 }, 053004 (2001).


\bibitem{AguilarSaavedra:2001rg}
  J.~A.~Aguilar-Saavedra {\it et al.}  [ECFA/DESY LC Physics Working Group],
  arXiv:hep-ph/0106315.

\bibitem{Heinemeyer:2007aw}
  S.~Heinemeyer, W.~Hollik, A.~M.~Weber and G.~Weiglein,
  arXiv:0711.0456 [hep-ph].

\bibitem{Erler:2008ek}
  J.~Erler and P.~Langacker,
  arXiv:0807.3023 [hep-ph].

\bibitem{Cirigliano:2007xi}
V.~Cirigliano and I.~Rosell,
  Phys.\ Rev.\ Lett.\  {\bf 99}, 231801 (2007)
  [arXiv:0707.3439 [hep-ph]].

\end{thebibliography}
\end{document}